\documentclass[aps, prb, superscriptaddress, citeautoscript, showpacs, twocolumn,nolongbibliography]{revtex4-2}
\usepackage{float}
\usepackage[pdftex]{graphicx}
\usepackage{epstopdf}
\usepackage{amsmath}
\usepackage{subfigure}
\usepackage{dcolumn}
\usepackage{enumitem}
\usepackage{amsfonts}

\newcommand{\gd}{GdV$_6$Sn$_6$}

\begin{document}

\title{Incommensurate Magnetic Order in the $\mathbb{Z}_2$ Kagome Metal GdV$_6$Sn$_6$}

\author{Zach Porter}
\affiliation{Materials Department, University of California, Santa Barbara, California 93106, USA}
\affiliation{Linac Coherent Light Source and Stanford Institute for Materials and Energy Sciences, SLAC National Accelerator Laboratory, Menlo Park, California, 94025, USA}

\author{Ganesh Pokharel}
\affiliation{Materials Department, University of California, Santa Barbara, California 93106, USA}

\author{Jong-Woo Kim}
\author{Phillip J. Ryan}
\affiliation{X-ray Science Division, Argonne National Laboratory, Argonne, Illinois 60439, USA}

\author{Stephen D. Wilson}
\email[email: ]{stephendwilson@ucsb.edu}
\affiliation{Materials Department, University of California, Santa Barbara, California 93106, USA}
\affiliation{California Nanosystems Institute, University of California, Santa Barbara, California 93106, USA}

\date{\today}

\begin{abstract}
We characterize the magnetic ground state of the topological kagome metal \gd~via resonant X-ray diffraction. Previous magnetoentropic studies of \gd~suggested the presence of a modulated magnetic order distinct from the ferromagnetism that is easily polarized by the application of a magnetic field. Diffraction data near the Gd-$L_2$ edge directly resolve a $c$-axis modulated spin structure order on the Gd sublattice with an incommensurate wave vector that evolves upon cooling toward a partial lock-in transition. While equal moment (spiral) and amplitude (sine) modulated spin states can not be unambiguously discerned from the scattering data, the overall phenomenology suggests an amplitude modulated state with moments predominantly oriented in the $ab$-plane. Comparisons to the ``double-flat" spiral state observed in Mn-based $R$Mn$_6$Sn$_6$ kagome compounds of the same structure type are discussed.
\end{abstract}
\maketitle

Itinerant kagome lattices have drawn considerable recent attention due to the possibility of realizing correlated topological states and unconventional forms of electronic order. Recent examples of materials range from $A$V$_3$Sb$_5$ compounds hosting unconventional charge density wave order and superconductivity \cite{PhysRevMaterials.5.034801, PhysRevLett.125.247002, Yin_2021, Jiang2021, Zhao2021}, to compounds such as Fe$_3$Sn$_2$ \cite{doi:10.1063/1.5088173, Ye2018} and Mn$_3$Sn \cite{PhysRevB.101.094404,Nakatsuji2015} that manifest large anomalous/topological Hall effects, to systems such as FeGe where charge order becomes intertwined with magnetic order \cite{Teng2022}. 

One versatile class of kagome compounds are materials of the form $R$V$_6$Sn$_6$ where $R$ is a rare earth ion that forms a triangular lattice network in close proximity to a nonmagnetic kagome network of V-site ions \cite{ROMAKA20118862}. Choice of the $R$-site allows for control of the character of the magnetic order proximitized with the nonmagnetic kagome network, which dominates the bands at the Fermi level. These compounds are also interesting comparators to their Mn-based $R$Mn$_6$Sn$_6$ cousins, where magnetic order directly forms within the kagome network \cite{doi:10.1126/sciadv.abe2680, GORBUNOV201247, Jifan_Hu_1995}.

Of particular interest is the impact of magnetic order on the Dirac points and saddle points native to the kagome network and very close to the Fermi level in $R$V$_6$Sn$_6$ compounds \cite{PhysRevB.104.235139, PhysRevLett.127.266401}. Strong spin-orbit coupling derived primarily from the heavy Sn-sites intrinsically gaps Dirac points and can create a tunable Chern gap with a large anomalous Hall effect \cite{Yin2020}, and while ferromagnetism can modify the magnitude of this gap, other spin states proximitized with the topological bands are also of interest. 

A number of recent studies have reported differing magnetic anisotropies manifest in $R$V$_6$Sn$_6$ via the choice of $R$-site \cite{PhysRevB.104.235139, PhysRevMaterials.6.083401, PhysRevB.106.115139, PhysRevMaterials.6.104202, huang2023anisotropic, PhysRevMaterials.6.105001}. Strongly uniaxial ferromagnetism was reported in TbV$_6$Sn$_6$ \cite{PhysRevMaterials.6.104202, PhysRevB.106.115139} and easy-plane anistotropy was reported in compounds such as (Sm,Er)V$_6$Sn$_6$ \cite{huang2023anisotropic, PhysRevMaterials.6.105001}.
Of particular note, nearly isotropic magnetic order was observed in spin-only $S=7/2$ GdV$_6$Sn$_6$ below 5 K \cite{PhysRevB.104.235139, doi:10.7566/JPSJ.90.124704}. Near the transition however, an unusual anomaly was observed in magnetoentropy maps, hinting that the zero field state was not trivially ferromagnetic. This raises the interesting possibility of realizing an incommensurate, modulated spin state interfaced with the conducting V-based kagome network, perhaps mimicking the modulation observed in magnetic kagome lattices of the same structure type (e.g. YMn$_6$Sn$_6$) and key to stabilizing the topological Hall effect in those materials \cite{PhysRevB.103.094413, doi:10.1126/sciadv.abe2680, PhysRevB.103.014416}. To address this possibility, momentum-resolved studies of the magnetic order in GdV$_6$Sn$_6$ are needed.

In this paper, we present low-temperature resonant elastic X-ray scattering (REXS) measurements characterizing the magnetic ground state of GdV$_6$Sn$_6$. Our data reveal that the zero-field magnetic ground state is an inter-layer modulated, incommensurate state and with the moments likely predominantly oriented within the kagome planes. The incommensurate ordering wavevector evolves upon cooling below the first T$_{AF1}=5.2(1)$ K ordering transition and develops an additional, coexisting commensurate component below T$_{AF2}=3.8(2)$ K, suggestive of a lower temperature lock-in transition. Our results demonstrate that an incommensurate magnetic state can be stabilized in $R$V$_6$Sn$_6$ via choice of an nearly isotropic $R$-site ion, and the appearance of a low-temperature commensurate harmonic suggests that the nature of the order is amplitude-modulated and driven by extended RKKY interactions. 

\begin{figure}[t]
\includegraphics[trim=10mm 156mm 10mm 10mm, clip,width=0.48\textwidth]{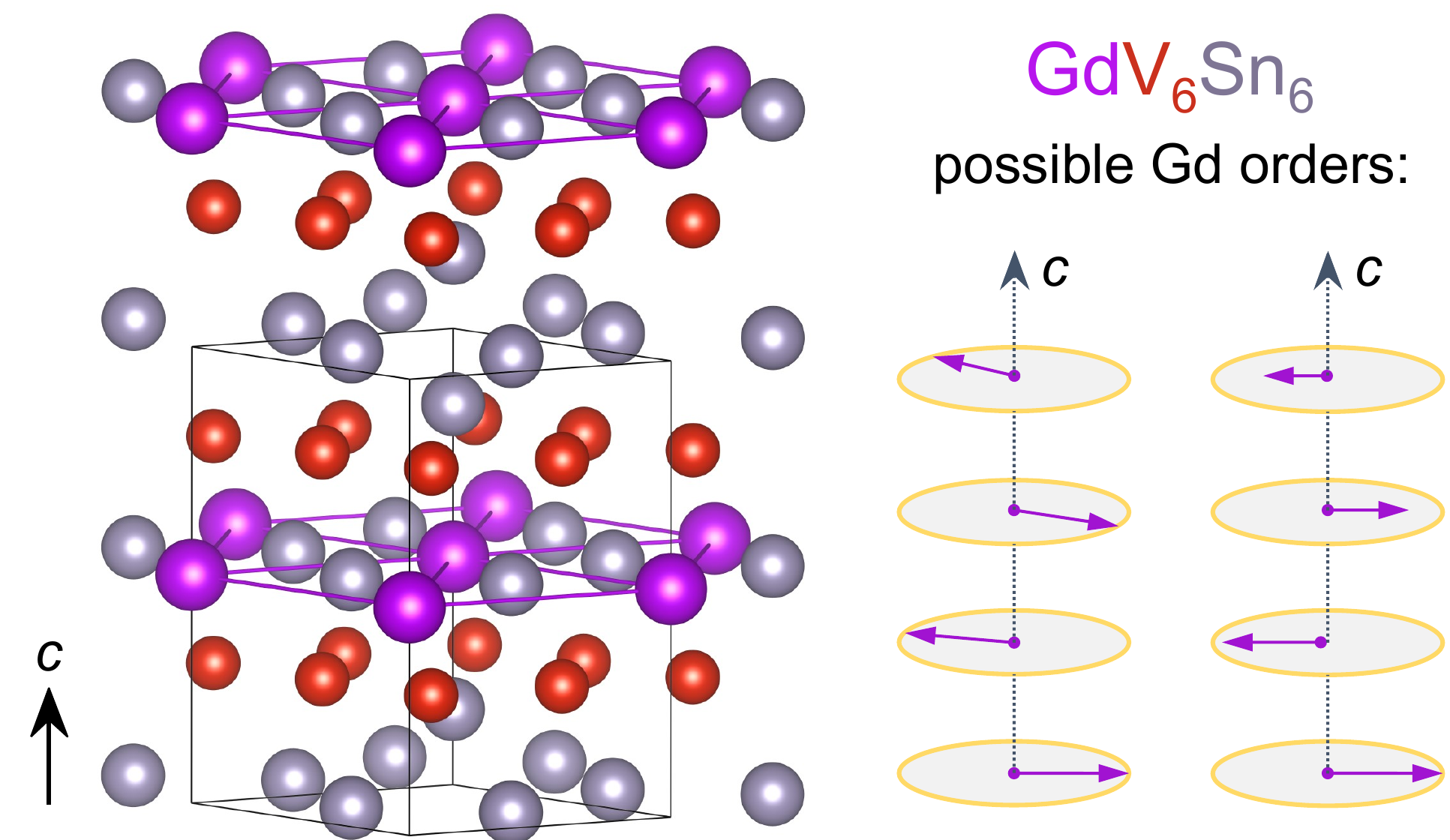}
\caption{Incommensurate magnetism in \gd. Gd forms hexagonal planes with one site per unit cell. Two possible Gd orders are shown: spiral and amplitude-modulated. The propagation direction is along $c$ but the moment orientation is uncertain.}
\label{fig:overview}
\end{figure}

For our studies, single crystal samples of \gd~were synthesized via a previously reported flux method \cite{PhysRevB.104.235139}. The unit cell is $P6/mmm$ with lattice parameters $a=b=5.532$ \AA~and $c=9.162$ \AA, and Gd is in Wyckoff site $1b$ at (0, 0, 0.5); see Figure \ref{fig:overview}. REXS measurements were performed on beamline 6-ID-B at the Advanced Photon Source at Argonne National Laboratory. The diffractometer endstation (Huber Psi-circle) was equipped with a Joule-Thomson stage displex cryostat capable of reaching a base temperature of 1.9 K. Measurements were performed near the Gd-$L_2$ resonance at $E_i=7.933$ keV using an area detector (Dectris Pilatus 100K) directly after the sample. Polarization analysis of the scattered beam was performed for select scans using a pyrolytic graphite (0,0,6) flat single crystal analyzer placed between the sample and a scintillation point detector. Magnetization was measured using a Quantum Design Magnetic Property Measurement System (MPMS3) with a single crystal attached to a quartz paddle with GE varnish. 

\begin{figure}[t]
\includegraphics[trim=10mm 14mm 55mm 10mm, clip,width=0.4\textwidth]{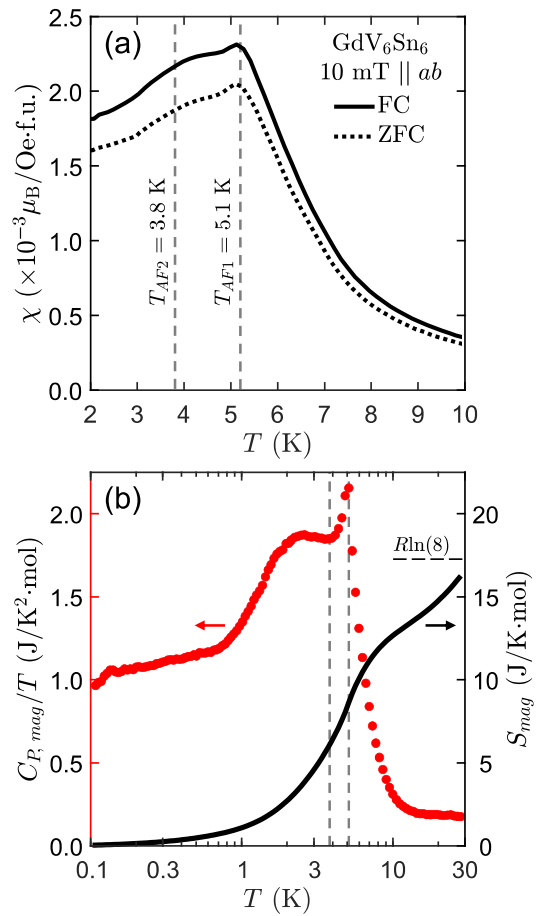}
\caption{Thermodynamics of \gd. Panel (a) shows the dc magnetic susceptibility $\chi$ measured on (zero) field cooling, denoted (Z)FC. Panel (b) shows the magnetic contribution to the heat capacity $C_{P,mag}/T$ (red dots) and entropy $S_{mag}$ (black line). Vertical dashed lines mark two phase transitions.}
\label{fig:maghc}
\end{figure}

Figure \ref{fig:maghc} shows thermodynamic measurements characterizing the onset of magnetic order in GdV$_6$Sn$_6$. Prior magnetoentropic studies identified the ground state as possibly noncollinear or modulated due to the presence of a low-field entropy barrier in the ordered state \cite{PhysRevB.104.235139}, since inflections in the temperature-dependent susceptibility evolve with increasing field \cite{doi:10.7566/JPSJ.90.124704}. Illustrating these low-field inflections, dc susceptibility $\chi$(\textit{T}) data are plotted in Figure \ref{fig:maghc}(a) with 10 mT applied within the basal $ab$-plane. Two features appear: the first is a cusp in the susceptibility at $T_{AF1}=5.2$ K, followed by a second, more subtle cusp near $T_{AF2}=3.8$ K. 

Heat capacity data also capture these two temperature scales as plotted in Figure \ref{fig:maghc}(b). Previously published $C_P$($T$) data for \gd~\cite{PhysRevB.104.235139} were analyzed to extract the magnetic entropy via removal of the phonon and charge contributions via a scaled subtraction of the nonmagnetic reference YV$_6$Sn$_6$ \cite{PhysRevB.43.13137}. The resulting magnetic $C_{P,mag}$($T$)/$T$ data are plotted along with the integrated entropy $S_{mag}$($T$). Both $T_{AF1}$ and $T_{AF2}$ can be identified in $C_{P,mag}$($T$)/$T$ via the peak and inflection point marked as dashed lines in Figure 2 (b). We note here that the Schottky anomaly expected due to the mean-field splitting of the $J=7/2$ ground state multiplet is expected to occur at $T_{AF1} / 4 \approx 1.3$ K and is likely hidden in the shoulder of the broad magnetic entropy peak. There is substantial entropy that extends to temperatures far above the initial ordering temperature, and the entropy integration up to 30 K nearly reaches the expected $R$ln(8); however uncertainties in the lattice subtraction and an imperfect lattice standard begin to dominate at these temperatures. Notably, there is substantial entropy that continues to be released at temperatures far below $T_{AF1}$, and as we will show next, scattering data reflect this via a continued staging of wave vectors as the low temperature limit is approached.

\begin{figure}
\includegraphics[trim=10mm 11mm 68mm 10mm, clip,width=0.364\textwidth]{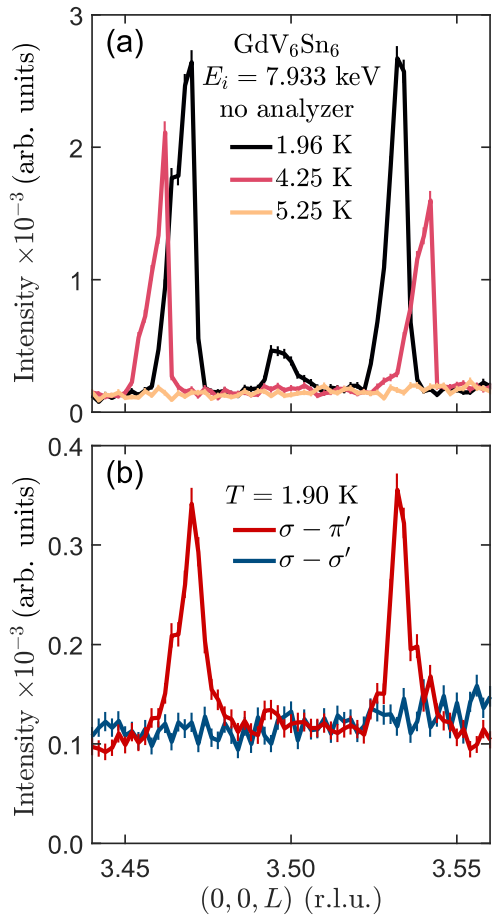}
\caption{Overview of the resonant X-ray diffraction. Both panels show scans along (0,0,$L$) at the Gd-$L_2$ resonance. Panel (a) shows scans at several temperatures without an analyzer crystal. Panel (b) shows scans using an analyzer crystal in the $\sigma-\pi'$ and $\sigma-\sigma'$ polarization channels.}
\label{fig:RXD}
\end{figure}

To explore the origins of these two features in $\chi$($T$) and $C_{P,mag}$($T$)/$T$ data, resonant X-ray scattering data were collected near the Gd-$L_2$ absorption edge. Magnetic superlattice reflections appear with a \textbf{q}=(0, 0, $l$) wave vector below the ordering temperatures $T_{AF1}$ and $T_{AF2}$ in a manner which mirrors the staged transition observed in $\chi$(\textit{T}). To further illustrate this staging, $L$-scans were performed about the \textbf{Q}=(0, 0, 3.5) position in three temperature regimes: $T>T_{AF1}$, $T_{AF2}>T>T_{AF1}$, and $T_{AF1}>T$. The resulting data are plotted in Figure \ref{fig:RXD}(a).

 In Figure \ref{fig:RXD}(a), upon cooling below $T_{AF1}$ an incommensurate set of reflections appear near \textbf{Q}=(0, 0, 3.47) and \textbf{Q}=(0, 0, 3.53) with \textbf{q$_{IC}$} ${\approx}$ $(0, 0, 0.47)$ below 5.2(1) K. Upon further cooling below $T_{AF2}$, an additional, weaker reflection appears at the commensurate \textbf{Q}=(0, 0, 3.50) position with \textbf{q$_{C}$}${=} (0, 0, 0.50)$ below 3.8(2) K and coexists with the incommensurate satellites. Both sets of peaks are long-range ordered, with widths of 0.007 r.l.u. along $L$ and 0.003 r.l.u. along $K$ that mirror those of nearby structural Bragg peaks. This indicates minimum correlation lengths of 0.2 microns that are constrained by the crystallinity of the sample.
 
 To further probe the origin of the incommensurate satellite peaks, polarization analysis was performed to separate the magnetic $\sigma-\pi'$ scattering channel from the nonmagnetic $\sigma-\sigma'$ channel. The results are plotted in Figure \ref{fig:RXD}(b), where the low temperature incommensurate peaks appear only in the $\sigma-\pi'$ channel, confirming their magnetic origin. The weaker commensurate peak is not clearly distinguishable in either polarization channel below $T_{AF2}$. However, the analyzer greatly diminishes signal-to-noise for these measurements. Perhaps the enhancement near \textbf{Q}=(0, 0, 3.50) in the $\sigma-\pi'$ channel suggests the peak is contributing. As will be shown next, the commensurate peak's temperature dependence correlates to the lower temperature anomaly in $\chi$(\textit{T}) and implies it also has a magnetic origin.

\begin{figure}
\includegraphics[trim=9mm 11mm 24.5mm 9mm, clip,width=0.48\textwidth]{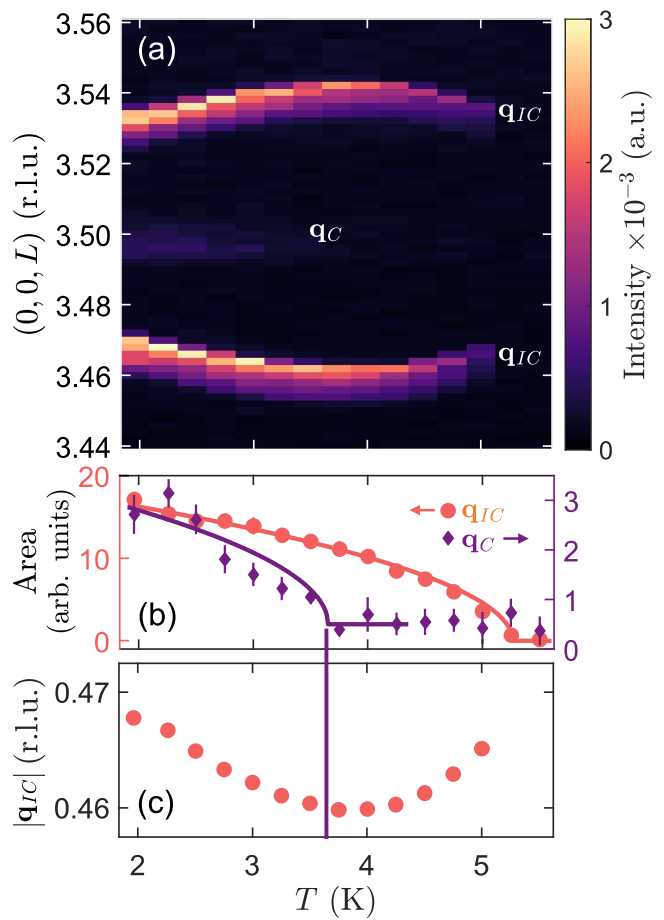}
\caption{Temperature dependence of the resonant X-ray diffraction. Panel (a) is a false color map of the scattering near (0, 0, 3.5). Three features are visible: two incommensurate reflections, and one half-integer commensurate reflection. Panels (b,c) are Gaussian fit results to these data; \textbf{q$_{IC}$} results are the weighted mean of both peaks. Panel (b) shows the area (integrated intensity) of \textbf{q$_{IC}$} (orange dots) and \textbf{q$_{C}$} (purple diamonds). Lines in (b) are guides to the eye of the form $\sqrt{1-T/T_{AF}}$. Panel (c) shows the wave vector $|$\textbf{q$_{IC}$}$|$ as a function of temperature.}
\label{fig:OP}
\end{figure}

The detailed temperature dependence of both the \textbf{q$_{IC}$} and \textbf{q$_{C}$} peaks are plotted in Figure \ref{fig:OP}(a). Peaks were parameterized via line scans at each temperature, and fit with Gaussian line shapes that quantify the peak area and position upon cooling toward base temperature. The resulting thermal evolution of both commensurate and incommensurate order parameters was well as the incommensurate wave vector \textbf{q$_{IC}$} are shown in Figures \ref{fig:OP}(b,c). 

Upon entering the incommensurate phase below $T_{AF1}$, \textbf{q$_{IC}$} begins to decrease with continued cooling and eventually reaches a minimum near $T_{AF2}$. Upon further cooling below $T_{AF2}$, weak scattering weight appears at the commensurate \textbf{q$_{C}$} position and the wavevector \textbf{q$_{IC}$} of the incommensurate peaks begins to \textit{increase}. This trend continues down to the lowest temperature probed ($\approx 2$ K). This interplay suggests that the discommensuration of the \textbf{q$_{IC}$} order is tied to the emergence of the \textbf{q$_{C}$} order parameter, consistent with a partial lock-in transition.



 In the susceptibility $\chi$(\textit{T}), the slight easy-plane anisotropy combined with the fact that both $T_{AF1}$ and $T_{AF2}$ appear only under an in-plane field suggest that the moments are predominantly in-plane. Magnetic scattering surveyed in multiple zones reveals that the scattering is strongest in $c$-axis aligned \textbf{Q}=(0, 0, $L$) type positions, also consistent with an in-plane component \footnote{See the Supplemental Material at [] for additional X-ray scattering and susceptibility measurements.}. However, in order to unambiguously determine the moment orientation, future measurements such as azimuthal scans at multiple wave vectors will be required. 

The incommensurate state observed in our REXS measurements is commonly indicative of either a flat spiral state or an amplitude-modulated state as illustrated by cartoons in Fig. 1. A flat spiral is reminiscent of the ``double flat" spiral states reported in Mn-based $R$Mn$_6$Sn$_6$ compounds \cite{ROSENFELD20081898, VENTURINI1996102}. A key distinction, however, is that the magnetic order in compounds such as YMn$_6$Sn$_6$ and LuMn$_6$Sn$_6$ derives from their Mn-based bilayer kagome networks, which na\"{i}vely host competing interlayer exchange interactions. Meanwhile magnetic order in $R$V$_6$Sn$_6$ compounds is derived from their triangular lattice lanthanide networks, where extended (beyond nearest neighbor) interlayer exchange interactions are much less likely. This suggests that RKKY interactions play a dominant role in the stabilization of the modulated magnetic state.

A further distinction is the development of a commensurate harmonic \textbf{q$_C$} at \textit{low temperature} in GdV$_6$Sn$_6$ versus the high temperature commensurate harmonic in (Y, Lu)Mn$_6$Sn$_6$ that splits into incommensurate wavelengths upon cooling \cite{VENTURINI1996102, PhysRevB.103.094413}. The onset of the \textbf{q$_C$} harmonic upon cooling below $T_{AF2}$ can be envisioned as the local formation of phase-locked modulations of antiferromagnetic planes that propagate within the longer wavelength, incommensurate state. The fact that this commensurate harmonic strengthens upon cooling toward the ground state points to the formation of a partial lock-in transition \cite{B_Penc_1999}---one which may develop further at lower temperatures. This transition toward an equal moment state at low temperature is suggestive of an amplitude modulated character to the incommensurate order. The substantial entropy remaining far below the onset of the incommensurate $T_{AF1}$ \cite{PhysRevB.43.13145, PhysRevB.64.014402} and the associated minimization of entropy achieved as the \textbf{q$_C$} harmonic develops are also consistent with an amplitude modulated nature to the transition. 

While the incommensurate state that forms below $T_{AF1}$ mimics those of (Y,Lu)Mn$_6$Sn$_6$, its origin should be distinct from the physics proposed in those compounds. The evolution of \textbf{q$_{IC}$} upon cooling below $T_{AF1}$ likely arises from entropic factors associated with an amplitude modulated state rather than lattice effects associated with unit cell changes tuning the balance of interlayer exchange interactions in a flat spiral picture. The incommensurate modulation wavevector \textbf{q$_{IC}$} is roughly consistent with nesting between the small Fermi surface pockets identified along the M-L line of the Brillouin zone \cite{doi:10.7566/JPSJ.90.124704}, again evoking an RKKY mechanism for the modulated state. Though future measurements examining whether the ordered spin state is collinear or noncollinear will be required, our current data demonstrate a complex condensation of magnetic order in the topological kagome metal GdV$_6$Sn$_6$. This provides an interesting avenue for engineering new magnetic states in proximity to the kagome planes in $R$V$_6$Sn$_6$ compounds and for exploring their impact on the topological band structures.


\begin{acknowledgments}
This work was supported by the National Science Foundation (NSF) through Enabling Quantum Leap: Convergent Accelerated Discovery Foundries for Quantum Materials Science, Engineering and Information (Q-AMASE-i): Quantum Foundry at UC Santa Barbara (Grant No. DMR-1906325). S.D.W. and Z.P. acknowledge support from NSF Grant No. DMR-1905801. This research made use of the shared facilities of the NSF Materials Research Science and Engineering Center at UC Santa Barbara, Grant No. DMR-1720256.  Z.P. acknowledges additional support from the U.S. Department of Energy, Office of Science, Basic Energy Sciences, Materials Sciences and Engineering Division, under Contract No. DE-AC02-76SF00515. This research used resources of the Advanced Photon Source, a U.S. Department of Energy (DOE) Office of Science User Facility operated for the DOE Office of Science by Argonne National Laboratory under Contract No. DEAC02-06CH11357.
\end{acknowledgments}


\bibliography{Gd166bib}

 \newcommand{\noop}[1]{}
\begin{thebibliography}{32}%
\makeatletter
\providecommand \@ifxundefined [1]{%
 \@ifx{#1\undefined}
}%
\providecommand \@ifnum [1]{%
 \ifnum #1\expandafter \@firstoftwo
 \else \expandafter \@secondoftwo
 \fi
}%
\providecommand \@ifx [1]{%
 \ifx #1\expandafter \@firstoftwo
 \else \expandafter \@secondoftwo
 \fi
}%
\providecommand \natexlab [1]{#1}%
\providecommand \enquote  [1]{``#1''}%
\providecommand \bibnamefont  [1]{#1}%
\providecommand \bibfnamefont [1]{#1}%
\providecommand \citenamefont [1]{#1}%
\providecommand \href@noop [0]{\@secondoftwo}%
\providecommand \href [0]{\begingroup \@sanitize@url \@href}%
\providecommand \@href[1]{\@@startlink{#1}\@@href}%
\providecommand \@@href[1]{\endgroup#1\@@endlink}%
\providecommand \@sanitize@url [0]{\catcode `\\12\catcode `\$12\catcode
  `\&12\catcode `\#12\catcode `\^12\catcode `\_12\catcode `\%12\relax}%
\providecommand \@@startlink[1]{}%
\providecommand \@@endlink[0]{}%
\providecommand \url  [0]{\begingroup\@sanitize@url \@url }%
\providecommand \@url [1]{\endgroup\@href {#1}{\urlprefix }}%
\providecommand \urlprefix  [0]{URL }%
\providecommand \Eprint [0]{\href }%
\providecommand \doibase [0]{https://doi.org/}%
\providecommand \selectlanguage [0]{\@gobble}%
\providecommand \bibinfo  [0]{\@secondoftwo}%
\providecommand \bibfield  [0]{\@secondoftwo}%
\providecommand \translation [1]{[#1]}%
\providecommand \BibitemOpen [0]{}%
\providecommand \bibitemStop [0]{}%
\providecommand \bibitemNoStop [0]{.\EOS\space}%
\providecommand \EOS [0]{\spacefactor3000\relax}%
\providecommand \BibitemShut  [1]{\csname bibitem#1\endcsname}%
\let\auto@bib@innerbib\@empty
\bibitem [{\citenamefont {Ortiz}\ \emph {et~al.}(2021)\citenamefont {Ortiz},
  \citenamefont {Sarte}, \citenamefont {Kenney}, \citenamefont {Graf},
  \citenamefont {Teicher}, \citenamefont {Seshadri},\ and\ \citenamefont
  {Wilson}}]{PhysRevMaterials.5.034801}%
  \BibitemOpen
  \bibfield  {author} {\bibinfo {author} {\bibfnamefont {B.~R.}\ \bibnamefont
  {Ortiz}}, \bibinfo {author} {\bibfnamefont {P.~M.}\ \bibnamefont {Sarte}},
  \bibinfo {author} {\bibfnamefont {E.~M.}\ \bibnamefont {Kenney}}, \bibinfo
  {author} {\bibfnamefont {M.~J.}\ \bibnamefont {Graf}}, \bibinfo {author}
  {\bibfnamefont {S.~M.~L.}\ \bibnamefont {Teicher}}, \bibinfo {author}
  {\bibfnamefont {R.}~\bibnamefont {Seshadri}},\ and\ \bibinfo {author}
  {\bibfnamefont {S.~D.}\ \bibnamefont {Wilson}},\ }\href
  {https://doi.org/10.1103/PhysRevMaterials.5.034801} {\bibfield  {journal}
  {\bibinfo  {journal} {Phys. Rev. Mater.}\ }\textbf {\bibinfo {volume} {5}},\
  \bibinfo {pages} {034801} (\bibinfo {year} {2021})}\BibitemShut {NoStop}%
\bibitem [{\citenamefont {Ortiz}\ \emph {et~al.}(2020)\citenamefont {Ortiz},
  \citenamefont {Teicher}, \citenamefont {Hu}, \citenamefont {Zuo},
  \citenamefont {Sarte}, \citenamefont {Schueller}, \citenamefont {Abeykoon},
  \citenamefont {Krogstad}, \citenamefont {Rosenkranz}, \citenamefont {Osborn},
  \citenamefont {Seshadri}, \citenamefont {Balents}, \citenamefont {He},\ and\
  \citenamefont {Wilson}}]{PhysRevLett.125.247002}%
  \BibitemOpen
  \bibfield  {author} {\bibinfo {author} {\bibfnamefont {B.~R.}\ \bibnamefont
  {Ortiz}}, \bibinfo {author} {\bibfnamefont {S.~M.~L.}\ \bibnamefont
  {Teicher}}, \bibinfo {author} {\bibfnamefont {Y.}~\bibnamefont {Hu}},
  \bibinfo {author} {\bibfnamefont {J.~L.}\ \bibnamefont {Zuo}}, \bibinfo
  {author} {\bibfnamefont {P.~M.}\ \bibnamefont {Sarte}}, \bibinfo {author}
  {\bibfnamefont {E.~C.}\ \bibnamefont {Schueller}}, \bibinfo {author}
  {\bibfnamefont {A.~M.~M.}\ \bibnamefont {Abeykoon}}, \bibinfo {author}
  {\bibfnamefont {M.~J.}\ \bibnamefont {Krogstad}}, \bibinfo {author}
  {\bibfnamefont {S.}~\bibnamefont {Rosenkranz}}, \bibinfo {author}
  {\bibfnamefont {R.}~\bibnamefont {Osborn}}, \bibinfo {author} {\bibfnamefont
  {R.}~\bibnamefont {Seshadri}}, \bibinfo {author} {\bibfnamefont
  {L.}~\bibnamefont {Balents}}, \bibinfo {author} {\bibfnamefont
  {J.}~\bibnamefont {He}},\ and\ \bibinfo {author} {\bibfnamefont {S.~D.}\
  \bibnamefont {Wilson}},\ }\href
  {https://doi.org/10.1103/PhysRevLett.125.247002} {\bibfield  {journal}
  {\bibinfo  {journal} {Phys. Rev. Lett.}\ }\textbf {\bibinfo {volume} {125}},\
  \bibinfo {pages} {247002} (\bibinfo {year} {2020})}\BibitemShut {NoStop}%
\bibitem [{\citenamefont {Yin}\ \emph {et~al.}(2021)\citenamefont {Yin},
  \citenamefont {Tu}, \citenamefont {Gong}, \citenamefont {Fu}, \citenamefont
  {Yan},\ and\ \citenamefont {Lei}}]{Yin_2021}%
  \BibitemOpen
  \bibfield  {author} {\bibinfo {author} {\bibfnamefont {Q.}~\bibnamefont
  {Yin}}, \bibinfo {author} {\bibfnamefont {Z.}~\bibnamefont {Tu}}, \bibinfo
  {author} {\bibfnamefont {C.}~\bibnamefont {Gong}}, \bibinfo {author}
  {\bibfnamefont {Y.}~\bibnamefont {Fu}}, \bibinfo {author} {\bibfnamefont
  {S.}~\bibnamefont {Yan}},\ and\ \bibinfo {author} {\bibfnamefont
  {H.}~\bibnamefont {Lei}},\ }\href
  {https://doi.org/10.1088/0256-307X/38/3/037403} {\bibfield  {journal}
  {\bibinfo  {journal} {Chinese Physics Letters}\ }\textbf {\bibinfo {volume}
  {38}},\ \bibinfo {pages} {037403} (\bibinfo {year} {2021})}\BibitemShut
  {NoStop}%
\bibitem [{\citenamefont {Jiang}\ \emph {et~al.}(2021)\citenamefont {Jiang},
  \citenamefont {Yin}, \citenamefont {Denner}, \citenamefont {Shumiya},
  \citenamefont {Ortiz}, \citenamefont {Xu}, \citenamefont {Guguchia},
  \citenamefont {He}, \citenamefont {Hossain}, \citenamefont {Liu},
  \citenamefont {Ruff}, \citenamefont {Kautzsch}, \citenamefont {Zhang},
  \citenamefont {Chang}, \citenamefont {Belopolski}, \citenamefont {Zhang},
  \citenamefont {Cochran}, \citenamefont {Multer}, \citenamefont {Litskevich},
  \citenamefont {Cheng}, \citenamefont {Yang}, \citenamefont {Wang},
  \citenamefont {Thomale}, \citenamefont {Neupert}, \citenamefont {Wilson},\
  and\ \citenamefont {Hasan}}]{Jiang2021}%
  \BibitemOpen
  \bibfield  {author} {\bibinfo {author} {\bibfnamefont {Y.-X.}\ \bibnamefont
  {Jiang}}, \bibinfo {author} {\bibfnamefont {J.-X.}\ \bibnamefont {Yin}},
  \bibinfo {author} {\bibfnamefont {M.~M.}\ \bibnamefont {Denner}}, \bibinfo
  {author} {\bibfnamefont {N.}~\bibnamefont {Shumiya}}, \bibinfo {author}
  {\bibfnamefont {B.~R.}\ \bibnamefont {Ortiz}}, \bibinfo {author}
  {\bibfnamefont {G.}~\bibnamefont {Xu}}, \bibinfo {author} {\bibfnamefont
  {Z.}~\bibnamefont {Guguchia}}, \bibinfo {author} {\bibfnamefont
  {J.}~\bibnamefont {He}}, \bibinfo {author} {\bibfnamefont {M.~S.}\
  \bibnamefont {Hossain}}, \bibinfo {author} {\bibfnamefont {X.}~\bibnamefont
  {Liu}}, \bibinfo {author} {\bibfnamefont {J.}~\bibnamefont {Ruff}}, \bibinfo
  {author} {\bibfnamefont {L.}~\bibnamefont {Kautzsch}}, \bibinfo {author}
  {\bibfnamefont {S.~S.}\ \bibnamefont {Zhang}}, \bibinfo {author}
  {\bibfnamefont {G.}~\bibnamefont {Chang}}, \bibinfo {author} {\bibfnamefont
  {I.}~\bibnamefont {Belopolski}}, \bibinfo {author} {\bibfnamefont
  {Q.}~\bibnamefont {Zhang}}, \bibinfo {author} {\bibfnamefont {T.~A.}\
  \bibnamefont {Cochran}}, \bibinfo {author} {\bibfnamefont {D.}~\bibnamefont
  {Multer}}, \bibinfo {author} {\bibfnamefont {M.}~\bibnamefont {Litskevich}},
  \bibinfo {author} {\bibfnamefont {Z.-J.}\ \bibnamefont {Cheng}}, \bibinfo
  {author} {\bibfnamefont {X.~P.}\ \bibnamefont {Yang}}, \bibinfo {author}
  {\bibfnamefont {Z.}~\bibnamefont {Wang}}, \bibinfo {author} {\bibfnamefont
  {R.}~\bibnamefont {Thomale}}, \bibinfo {author} {\bibfnamefont
  {T.}~\bibnamefont {Neupert}}, \bibinfo {author} {\bibfnamefont {S.~D.}\
  \bibnamefont {Wilson}},\ and\ \bibinfo {author} {\bibfnamefont {M.~Z.}\
  \bibnamefont {Hasan}},\ }\href {https://doi.org/10.1038/s41563-021-01034-y}
  {\bibfield  {journal} {\bibinfo  {journal} {Nature Materials}\ }\textbf
  {\bibinfo {volume} {20}},\ \bibinfo {pages} {1353} (\bibinfo {year}
  {2021})}\BibitemShut {NoStop}%
\bibitem [{\citenamefont {Zhao}\ \emph {et~al.}(2021)\citenamefont {Zhao},
  \citenamefont {Li}, \citenamefont {Ortiz}, \citenamefont {Teicher},
  \citenamefont {Park}, \citenamefont {Ye}, \citenamefont {Wang}, \citenamefont
  {Balents}, \citenamefont {Wilson},\ and\ \citenamefont
  {Zeljkovic}}]{Zhao2021}%
  \BibitemOpen
  \bibfield  {author} {\bibinfo {author} {\bibfnamefont {H.}~\bibnamefont
  {Zhao}}, \bibinfo {author} {\bibfnamefont {H.}~\bibnamefont {Li}}, \bibinfo
  {author} {\bibfnamefont {B.~R.}\ \bibnamefont {Ortiz}}, \bibinfo {author}
  {\bibfnamefont {S.~M.~L.}\ \bibnamefont {Teicher}}, \bibinfo {author}
  {\bibfnamefont {T.}~\bibnamefont {Park}}, \bibinfo {author} {\bibfnamefont
  {M.}~\bibnamefont {Ye}}, \bibinfo {author} {\bibfnamefont {Z.}~\bibnamefont
  {Wang}}, \bibinfo {author} {\bibfnamefont {L.}~\bibnamefont {Balents}},
  \bibinfo {author} {\bibfnamefont {S.~D.}\ \bibnamefont {Wilson}},\ and\
  \bibinfo {author} {\bibfnamefont {I.}~\bibnamefont {Zeljkovic}},\ }\href
  {https://doi.org/10.1038/s41586-021-03946-w} {\bibfield  {journal} {\bibinfo
  {journal} {Nature}\ }\textbf {\bibinfo {volume} {599}},\ \bibinfo {pages}
  {216} (\bibinfo {year} {2021})}\BibitemShut {NoStop}%
\bibitem [{\citenamefont {Li}\ \emph {et~al.}(2019)\citenamefont {Li},
  \citenamefont {Ding}, \citenamefont {Chen}, \citenamefont {Li}, \citenamefont
  {Hou}, \citenamefont {Liu}, \citenamefont {Zhang}, \citenamefont {Xi},
  \citenamefont {Wu},\ and\ \citenamefont {Wang}}]{doi:10.1063/1.5088173}%
  \BibitemOpen
  \bibfield  {author} {\bibinfo {author} {\bibfnamefont {H.}~\bibnamefont
  {Li}}, \bibinfo {author} {\bibfnamefont {B.}~\bibnamefont {Ding}}, \bibinfo
  {author} {\bibfnamefont {J.}~\bibnamefont {Chen}}, \bibinfo {author}
  {\bibfnamefont {Z.}~\bibnamefont {Li}}, \bibinfo {author} {\bibfnamefont
  {Z.}~\bibnamefont {Hou}}, \bibinfo {author} {\bibfnamefont {E.}~\bibnamefont
  {Liu}}, \bibinfo {author} {\bibfnamefont {H.}~\bibnamefont {Zhang}}, \bibinfo
  {author} {\bibfnamefont {X.}~\bibnamefont {Xi}}, \bibinfo {author}
  {\bibfnamefont {G.}~\bibnamefont {Wu}},\ and\ \bibinfo {author}
  {\bibfnamefont {W.}~\bibnamefont {Wang}},\ }\href
  {https://doi.org/10.1063/1.5088173} {\bibfield  {journal} {\bibinfo
  {journal} {Applied Physics Letters}\ }\textbf {\bibinfo {volume} {114}},\
  \bibinfo {pages} {192408} (\bibinfo {year} {2019})},\ \Eprint
  {https://arxiv.org/abs/https://doi.org/10.1063/1.5088173}
  {https://doi.org/10.1063/1.5088173} \BibitemShut {NoStop}%
\bibitem [{\citenamefont {Ye}\ \emph {et~al.}(2018)\citenamefont {Ye},
  \citenamefont {Kang}, \citenamefont {Liu}, \citenamefont {von Cube},
  \citenamefont {Wicker}, \citenamefont {Suzuki}, \citenamefont {Jozwiak},
  \citenamefont {Bostwick}, \citenamefont {Rotenberg}, \citenamefont {Bell},
  \citenamefont {Fu}, \citenamefont {Comin},\ and\ \citenamefont
  {Checkelsky}}]{Ye2018}%
  \BibitemOpen
  \bibfield  {author} {\bibinfo {author} {\bibfnamefont {L.}~\bibnamefont
  {Ye}}, \bibinfo {author} {\bibfnamefont {M.}~\bibnamefont {Kang}}, \bibinfo
  {author} {\bibfnamefont {J.}~\bibnamefont {Liu}}, \bibinfo {author}
  {\bibfnamefont {F.}~\bibnamefont {von Cube}}, \bibinfo {author}
  {\bibfnamefont {C.~R.}\ \bibnamefont {Wicker}}, \bibinfo {author}
  {\bibfnamefont {T.}~\bibnamefont {Suzuki}}, \bibinfo {author} {\bibfnamefont
  {C.}~\bibnamefont {Jozwiak}}, \bibinfo {author} {\bibfnamefont
  {A.}~\bibnamefont {Bostwick}}, \bibinfo {author} {\bibfnamefont
  {E.}~\bibnamefont {Rotenberg}}, \bibinfo {author} {\bibfnamefont {D.~C.}\
  \bibnamefont {Bell}}, \bibinfo {author} {\bibfnamefont {L.}~\bibnamefont
  {Fu}}, \bibinfo {author} {\bibfnamefont {R.}~\bibnamefont {Comin}},\ and\
  \bibinfo {author} {\bibfnamefont {J.~G.}\ \bibnamefont {Checkelsky}},\ }\href
  {https://doi.org/10.1038/nature25987} {\bibfield  {journal} {\bibinfo
  {journal} {Nature}\ }\textbf {\bibinfo {volume} {555}},\ \bibinfo {pages}
  {638} (\bibinfo {year} {2018})}\BibitemShut {NoStop}%
\bibitem [{\citenamefont {Taylor}\ \emph {et~al.}(2020)\citenamefont {Taylor},
  \citenamefont {Markou}, \citenamefont {Lesne}, \citenamefont {Sivakumar},
  \citenamefont {Luo}, \citenamefont {Radu}, \citenamefont {Werner},
  \citenamefont {Felser},\ and\ \citenamefont {Parkin}}]{PhysRevB.101.094404}%
  \BibitemOpen
  \bibfield  {author} {\bibinfo {author} {\bibfnamefont {J.~M.}\ \bibnamefont
  {Taylor}}, \bibinfo {author} {\bibfnamefont {A.}~\bibnamefont {Markou}},
  \bibinfo {author} {\bibfnamefont {E.}~\bibnamefont {Lesne}}, \bibinfo
  {author} {\bibfnamefont {P.~K.}\ \bibnamefont {Sivakumar}}, \bibinfo {author}
  {\bibfnamefont {C.}~\bibnamefont {Luo}}, \bibinfo {author} {\bibfnamefont
  {F.}~\bibnamefont {Radu}}, \bibinfo {author} {\bibfnamefont {P.}~\bibnamefont
  {Werner}}, \bibinfo {author} {\bibfnamefont {C.}~\bibnamefont {Felser}},\
  and\ \bibinfo {author} {\bibfnamefont {S.~S.~P.}\ \bibnamefont {Parkin}},\
  }\href {https://doi.org/10.1103/PhysRevB.101.094404} {\bibfield  {journal}
  {\bibinfo  {journal} {Phys. Rev. B}\ }\textbf {\bibinfo {volume} {101}},\
  \bibinfo {pages} {094404} (\bibinfo {year} {2020})}\BibitemShut {NoStop}%
\bibitem [{\citenamefont {Nakatsuji}\ \emph {et~al.}(2015)\citenamefont
  {Nakatsuji}, \citenamefont {Kiyohara},\ and\ \citenamefont
  {Higo}}]{Nakatsuji2015}%
  \BibitemOpen
  \bibfield  {author} {\bibinfo {author} {\bibfnamefont {S.}~\bibnamefont
  {Nakatsuji}}, \bibinfo {author} {\bibfnamefont {N.}~\bibnamefont
  {Kiyohara}},\ and\ \bibinfo {author} {\bibfnamefont {T.}~\bibnamefont
  {Higo}},\ }\href {https://doi.org/10.1038/nature15723} {\bibfield  {journal}
  {\bibinfo  {journal} {Nature}\ }\textbf {\bibinfo {volume} {527}},\ \bibinfo
  {pages} {212} (\bibinfo {year} {2015})}\BibitemShut {NoStop}%
\bibitem [{\citenamefont {Teng}\ \emph {et~al.}(2022)\citenamefont {Teng},
  \citenamefont {Chen}, \citenamefont {Ye}, \citenamefont {Rosenberg},
  \citenamefont {Liu}, \citenamefont {Yin}, \citenamefont {Jiang},
  \citenamefont {Oh}, \citenamefont {Hasan}, \citenamefont {Neubauer},
  \citenamefont {Gao}, \citenamefont {Xie}, \citenamefont {Hashimoto},
  \citenamefont {Lu}, \citenamefont {Jozwiak}, \citenamefont {Bostwick},
  \citenamefont {Rotenberg}, \citenamefont {Birgeneau}, \citenamefont {Chu},
  \citenamefont {Yi},\ and\ \citenamefont {Dai}}]{Teng2022}%
  \BibitemOpen
  \bibfield  {author} {\bibinfo {author} {\bibfnamefont {X.}~\bibnamefont
  {Teng}}, \bibinfo {author} {\bibfnamefont {L.}~\bibnamefont {Chen}}, \bibinfo
  {author} {\bibfnamefont {F.}~\bibnamefont {Ye}}, \bibinfo {author}
  {\bibfnamefont {E.}~\bibnamefont {Rosenberg}}, \bibinfo {author}
  {\bibfnamefont {Z.}~\bibnamefont {Liu}}, \bibinfo {author} {\bibfnamefont
  {J.-X.}\ \bibnamefont {Yin}}, \bibinfo {author} {\bibfnamefont {Y.-X.}\
  \bibnamefont {Jiang}}, \bibinfo {author} {\bibfnamefont {J.~S.}\ \bibnamefont
  {Oh}}, \bibinfo {author} {\bibfnamefont {M.~Z.}\ \bibnamefont {Hasan}},
  \bibinfo {author} {\bibfnamefont {K.~J.}\ \bibnamefont {Neubauer}}, \bibinfo
  {author} {\bibfnamefont {B.}~\bibnamefont {Gao}}, \bibinfo {author}
  {\bibfnamefont {Y.}~\bibnamefont {Xie}}, \bibinfo {author} {\bibfnamefont
  {M.}~\bibnamefont {Hashimoto}}, \bibinfo {author} {\bibfnamefont
  {D.}~\bibnamefont {Lu}}, \bibinfo {author} {\bibfnamefont {C.}~\bibnamefont
  {Jozwiak}}, \bibinfo {author} {\bibfnamefont {A.}~\bibnamefont {Bostwick}},
  \bibinfo {author} {\bibfnamefont {E.}~\bibnamefont {Rotenberg}}, \bibinfo
  {author} {\bibfnamefont {R.~J.}\ \bibnamefont {Birgeneau}}, \bibinfo {author}
  {\bibfnamefont {J.-H.}\ \bibnamefont {Chu}}, \bibinfo {author} {\bibfnamefont
  {M.}~\bibnamefont {Yi}},\ and\ \bibinfo {author} {\bibfnamefont
  {P.}~\bibnamefont {Dai}},\ }\href
  {https://doi.org/10.1038/s41586-022-05034-z} {\bibfield  {journal} {\bibinfo
  {journal} {Nature}\ }\textbf {\bibinfo {volume} {609}},\ \bibinfo {pages}
  {490} (\bibinfo {year} {2022})}\BibitemShut {NoStop}%
\bibitem [{\citenamefont {Romaka}\ \emph {et~al.}(2011)\citenamefont {Romaka},
  \citenamefont {Stadnyk}, \citenamefont {Romaka}, \citenamefont {Demchenko},
  \citenamefont {Stadnyshyn},\ and\ \citenamefont {Konyk}}]{ROMAKA20118862}%
  \BibitemOpen
  \bibfield  {author} {\bibinfo {author} {\bibfnamefont {L.}~\bibnamefont
  {Romaka}}, \bibinfo {author} {\bibfnamefont {Y.}~\bibnamefont {Stadnyk}},
  \bibinfo {author} {\bibfnamefont {V.}~\bibnamefont {Romaka}}, \bibinfo
  {author} {\bibfnamefont {P.}~\bibnamefont {Demchenko}}, \bibinfo {author}
  {\bibfnamefont {M.}~\bibnamefont {Stadnyshyn}},\ and\ \bibinfo {author}
  {\bibfnamefont {M.}~\bibnamefont {Konyk}},\ }\href
  {https://doi.org/https://doi.org/10.1016/j.jallcom.2011.06.095} {\bibfield
  {journal} {\bibinfo  {journal} {Journal of Alloys and Compounds}\ }\textbf
  {\bibinfo {volume} {509}},\ \bibinfo {pages} {8862} (\bibinfo {year}
  {2011})}\BibitemShut {NoStop}%
\bibitem [{\citenamefont {Ghimire}\ \emph {et~al.}(2020)\citenamefont
  {Ghimire}, \citenamefont {Dally}, \citenamefont {Poudel}, \citenamefont
  {Jones}, \citenamefont {Michel}, \citenamefont {Magar}, \citenamefont
  {Bleuel}, \citenamefont {McGuire}, \citenamefont {Jiang}, \citenamefont
  {Mitchell}, \citenamefont {Lynn},\ and\ \citenamefont
  {Mazin}}]{doi:10.1126/sciadv.abe2680}%
  \BibitemOpen
  \bibfield  {author} {\bibinfo {author} {\bibfnamefont {N.~J.}\ \bibnamefont
  {Ghimire}}, \bibinfo {author} {\bibfnamefont {R.~L.}\ \bibnamefont {Dally}},
  \bibinfo {author} {\bibfnamefont {L.}~\bibnamefont {Poudel}}, \bibinfo
  {author} {\bibfnamefont {D.~C.}\ \bibnamefont {Jones}}, \bibinfo {author}
  {\bibfnamefont {D.}~\bibnamefont {Michel}}, \bibinfo {author} {\bibfnamefont
  {N.~T.}\ \bibnamefont {Magar}}, \bibinfo {author} {\bibfnamefont
  {M.}~\bibnamefont {Bleuel}}, \bibinfo {author} {\bibfnamefont {M.~A.}\
  \bibnamefont {McGuire}}, \bibinfo {author} {\bibfnamefont {J.~S.}\
  \bibnamefont {Jiang}}, \bibinfo {author} {\bibfnamefont {J.~F.}\ \bibnamefont
  {Mitchell}}, \bibinfo {author} {\bibfnamefont {J.~W.}\ \bibnamefont {Lynn}},\
  and\ \bibinfo {author} {\bibfnamefont {I.~I.}\ \bibnamefont {Mazin}},\ }\href
  {https://doi.org/10.1126/sciadv.abe2680} {\bibfield  {journal} {\bibinfo
  {journal} {Science Advances}\ }\textbf {\bibinfo {volume} {6}},\ \bibinfo
  {pages} {eabe2680} (\bibinfo {year} {2020})},\ \Eprint
  {https://arxiv.org/abs/https://www.science.org/doi/pdf/10.1126/sciadv.abe2680}
  {https://www.science.org/doi/pdf/10.1126/sciadv.abe2680} \BibitemShut
  {NoStop}%
\bibitem [{\citenamefont {Gorbunov}\ \emph {et~al.}(2012)\citenamefont
  {Gorbunov}, \citenamefont {Kuz’min}, \citenamefont {Uhlířová},
  \citenamefont {Žáček}, \citenamefont {Richter}, \citenamefont {Skourski},\
  and\ \citenamefont {Andreev}}]{GORBUNOV201247}%
  \BibitemOpen
  \bibfield  {author} {\bibinfo {author} {\bibfnamefont {D.}~\bibnamefont
  {Gorbunov}}, \bibinfo {author} {\bibfnamefont {M.}~\bibnamefont {Kuz’min}},
  \bibinfo {author} {\bibfnamefont {K.}~\bibnamefont {Uhlířová}}, \bibinfo
  {author} {\bibfnamefont {M.}~\bibnamefont {Žáček}}, \bibinfo {author}
  {\bibfnamefont {M.}~\bibnamefont {Richter}}, \bibinfo {author} {\bibfnamefont
  {Y.}~\bibnamefont {Skourski}},\ and\ \bibinfo {author} {\bibfnamefont
  {A.}~\bibnamefont {Andreev}},\ }\href
  {https://doi.org/https://doi.org/10.1016/j.jallcom.2011.12.016} {\bibfield
  {journal} {\bibinfo  {journal} {Journal of Alloys and Compounds}\ }\textbf
  {\bibinfo {volume} {519}},\ \bibinfo {pages} {47} (\bibinfo {year}
  {2012})}\BibitemShut {NoStop}%
\bibitem [{\citenamefont {Hu}\ \emph {et~al.}(1995)\citenamefont {Hu},
  \citenamefont {Wang}, \citenamefont {Hu}, \citenamefont {Wang}, \citenamefont
  {Wang}, \citenamefont {Yang}, \citenamefont {Tang}, \citenamefont {Zhao},\
  and\ \citenamefont {Qin}}]{Jifan_Hu_1995}%
  \BibitemOpen
  \bibfield  {author} {\bibinfo {author} {\bibfnamefont {J.}~\bibnamefont
  {Hu}}, \bibinfo {author} {\bibfnamefont {K.-Y.}\ \bibnamefont {Wang}},
  \bibinfo {author} {\bibfnamefont {B.-P.}\ \bibnamefont {Hu}}, \bibinfo
  {author} {\bibfnamefont {Y.-Z.}\ \bibnamefont {Wang}}, \bibinfo {author}
  {\bibfnamefont {Z.}~\bibnamefont {Wang}}, \bibinfo {author} {\bibfnamefont
  {F.}~\bibnamefont {Yang}}, \bibinfo {author} {\bibfnamefont {N.}~\bibnamefont
  {Tang}}, \bibinfo {author} {\bibfnamefont {R.}~\bibnamefont {Zhao}},\ and\
  \bibinfo {author} {\bibfnamefont {W.}~\bibnamefont {Qin}},\ }\href
  {https://doi.org/10.1088/0953-8984/7/5/011} {\bibfield  {journal} {\bibinfo
  {journal} {Journal of Physics: Condensed Matter}\ }\textbf {\bibinfo {volume}
  {7}},\ \bibinfo {pages} {889} (\bibinfo {year} {1995})}\BibitemShut {NoStop}%
\bibitem [{\citenamefont {Pokharel}\ \emph {et~al.}(2021)\citenamefont
  {Pokharel}, \citenamefont {Teicher}, \citenamefont {Ortiz}, \citenamefont
  {Sarte}, \citenamefont {Wu}, \citenamefont {Peng}, \citenamefont {He},
  \citenamefont {Seshadri},\ and\ \citenamefont
  {Wilson}}]{PhysRevB.104.235139}%
  \BibitemOpen
  \bibfield  {author} {\bibinfo {author} {\bibfnamefont {G.}~\bibnamefont
  {Pokharel}}, \bibinfo {author} {\bibfnamefont {S.~M.~L.}\ \bibnamefont
  {Teicher}}, \bibinfo {author} {\bibfnamefont {B.~R.}\ \bibnamefont {Ortiz}},
  \bibinfo {author} {\bibfnamefont {P.~M.}\ \bibnamefont {Sarte}}, \bibinfo
  {author} {\bibfnamefont {G.}~\bibnamefont {Wu}}, \bibinfo {author}
  {\bibfnamefont {S.}~\bibnamefont {Peng}}, \bibinfo {author} {\bibfnamefont
  {J.}~\bibnamefont {He}}, \bibinfo {author} {\bibfnamefont {R.}~\bibnamefont
  {Seshadri}},\ and\ \bibinfo {author} {\bibfnamefont {S.~D.}\ \bibnamefont
  {Wilson}},\ }\href {https://doi.org/10.1103/PhysRevB.104.235139} {\bibfield
  {journal} {\bibinfo  {journal} {Phys. Rev. B}\ }\textbf {\bibinfo {volume}
  {104}},\ \bibinfo {pages} {235139} (\bibinfo {year} {2021})}\BibitemShut
  {NoStop}%
\bibitem [{\citenamefont {Peng}\ \emph {et~al.}(2021)\citenamefont {Peng},
  \citenamefont {Han}, \citenamefont {Pokharel}, \citenamefont {Shen},
  \citenamefont {Li}, \citenamefont {Hashimoto}, \citenamefont {Lu},
  \citenamefont {Ortiz}, \citenamefont {Luo}, \citenamefont {Li}, \citenamefont
  {Guo}, \citenamefont {Wang}, \citenamefont {Cui}, \citenamefont {Sun},
  \citenamefont {Qiao}, \citenamefont {Wilson},\ and\ \citenamefont
  {He}}]{PhysRevLett.127.266401}%
  \BibitemOpen
  \bibfield  {author} {\bibinfo {author} {\bibfnamefont {S.}~\bibnamefont
  {Peng}}, \bibinfo {author} {\bibfnamefont {Y.}~\bibnamefont {Han}}, \bibinfo
  {author} {\bibfnamefont {G.}~\bibnamefont {Pokharel}}, \bibinfo {author}
  {\bibfnamefont {J.}~\bibnamefont {Shen}}, \bibinfo {author} {\bibfnamefont
  {Z.}~\bibnamefont {Li}}, \bibinfo {author} {\bibfnamefont {M.}~\bibnamefont
  {Hashimoto}}, \bibinfo {author} {\bibfnamefont {D.}~\bibnamefont {Lu}},
  \bibinfo {author} {\bibfnamefont {B.~R.}\ \bibnamefont {Ortiz}}, \bibinfo
  {author} {\bibfnamefont {Y.}~\bibnamefont {Luo}}, \bibinfo {author}
  {\bibfnamefont {H.}~\bibnamefont {Li}}, \bibinfo {author} {\bibfnamefont
  {M.}~\bibnamefont {Guo}}, \bibinfo {author} {\bibfnamefont {B.}~\bibnamefont
  {Wang}}, \bibinfo {author} {\bibfnamefont {S.}~\bibnamefont {Cui}}, \bibinfo
  {author} {\bibfnamefont {Z.}~\bibnamefont {Sun}}, \bibinfo {author}
  {\bibfnamefont {Z.}~\bibnamefont {Qiao}}, \bibinfo {author} {\bibfnamefont
  {S.~D.}\ \bibnamefont {Wilson}},\ and\ \bibinfo {author} {\bibfnamefont
  {J.}~\bibnamefont {He}},\ }\href
  {https://doi.org/10.1103/PhysRevLett.127.266401} {\bibfield  {journal}
  {\bibinfo  {journal} {Phys. Rev. Lett.}\ }\textbf {\bibinfo {volume} {127}},\
  \bibinfo {pages} {266401} (\bibinfo {year} {2021})}\BibitemShut {NoStop}%
\bibitem [{\citenamefont {Yin}\ \emph {et~al.}(2020)\citenamefont {Yin},
  \citenamefont {Ma}, \citenamefont {Cochran}, \citenamefont {Xu},
  \citenamefont {Zhang}, \citenamefont {Tien}, \citenamefont {Shumiya},
  \citenamefont {Cheng}, \citenamefont {Jiang}, \citenamefont {Lian},
  \citenamefont {Song}, \citenamefont {Chang}, \citenamefont {Belopolski},
  \citenamefont {Multer}, \citenamefont {Litskevich}, \citenamefont {Cheng},
  \citenamefont {Yang}, \citenamefont {Swidler}, \citenamefont {Zhou},
  \citenamefont {Lin}, \citenamefont {Neupert}, \citenamefont {Wang},
  \citenamefont {Yao}, \citenamefont {Chang}, \citenamefont {Jia},\ and\
  \citenamefont {Zahid~Hasan}}]{Yin2020}%
  \BibitemOpen
  \bibfield  {author} {\bibinfo {author} {\bibfnamefont {J.-X.}\ \bibnamefont
  {Yin}}, \bibinfo {author} {\bibfnamefont {W.}~\bibnamefont {Ma}}, \bibinfo
  {author} {\bibfnamefont {T.~A.}\ \bibnamefont {Cochran}}, \bibinfo {author}
  {\bibfnamefont {X.}~\bibnamefont {Xu}}, \bibinfo {author} {\bibfnamefont
  {S.~S.}\ \bibnamefont {Zhang}}, \bibinfo {author} {\bibfnamefont {H.-J.}\
  \bibnamefont {Tien}}, \bibinfo {author} {\bibfnamefont {N.}~\bibnamefont
  {Shumiya}}, \bibinfo {author} {\bibfnamefont {G.}~\bibnamefont {Cheng}},
  \bibinfo {author} {\bibfnamefont {K.}~\bibnamefont {Jiang}}, \bibinfo
  {author} {\bibfnamefont {B.}~\bibnamefont {Lian}}, \bibinfo {author}
  {\bibfnamefont {Z.}~\bibnamefont {Song}}, \bibinfo {author} {\bibfnamefont
  {G.}~\bibnamefont {Chang}}, \bibinfo {author} {\bibfnamefont
  {I.}~\bibnamefont {Belopolski}}, \bibinfo {author} {\bibfnamefont
  {D.}~\bibnamefont {Multer}}, \bibinfo {author} {\bibfnamefont
  {M.}~\bibnamefont {Litskevich}}, \bibinfo {author} {\bibfnamefont {Z.-J.}\
  \bibnamefont {Cheng}}, \bibinfo {author} {\bibfnamefont {X.~P.}\ \bibnamefont
  {Yang}}, \bibinfo {author} {\bibfnamefont {B.}~\bibnamefont {Swidler}},
  \bibinfo {author} {\bibfnamefont {H.}~\bibnamefont {Zhou}}, \bibinfo {author}
  {\bibfnamefont {H.}~\bibnamefont {Lin}}, \bibinfo {author} {\bibfnamefont
  {T.}~\bibnamefont {Neupert}}, \bibinfo {author} {\bibfnamefont
  {Z.}~\bibnamefont {Wang}}, \bibinfo {author} {\bibfnamefont {N.}~\bibnamefont
  {Yao}}, \bibinfo {author} {\bibfnamefont {T.-R.}\ \bibnamefont {Chang}},
  \bibinfo {author} {\bibfnamefont {S.}~\bibnamefont {Jia}},\ and\ \bibinfo
  {author} {\bibfnamefont {M.}~\bibnamefont {Zahid~Hasan}},\ }\href
  {https://doi.org/10.1038/s41586-020-2482-7} {\bibfield  {journal} {\bibinfo
  {journal} {Nature}\ }\textbf {\bibinfo {volume} {583}},\ \bibinfo {pages}
  {533} (\bibinfo {year} {2020})}\BibitemShut {NoStop}%
\bibitem [{\citenamefont {Lee}\ and\ \citenamefont
  {Mun}(2022)}]{PhysRevMaterials.6.083401}%
  \BibitemOpen
  \bibfield  {author} {\bibinfo {author} {\bibfnamefont {J.}~\bibnamefont
  {Lee}}\ and\ \bibinfo {author} {\bibfnamefont {E.}~\bibnamefont {Mun}},\
  }\href {https://doi.org/10.1103/PhysRevMaterials.6.083401} {\bibfield
  {journal} {\bibinfo  {journal} {Phys. Rev. Mater.}\ }\textbf {\bibinfo
  {volume} {6}},\ \bibinfo {pages} {083401} (\bibinfo {year}
  {2022})}\BibitemShut {NoStop}%
\bibitem [{\citenamefont {Rosenberg}\ \emph {et~al.}(2022)\citenamefont
  {Rosenberg}, \citenamefont {DeStefano}, \citenamefont {Guo}, \citenamefont
  {Oh}, \citenamefont {Hashimoto}, \citenamefont {Lu}, \citenamefont
  {Birgeneau}, \citenamefont {Lee}, \citenamefont {Ke}, \citenamefont {Yi},\
  and\ \citenamefont {Chu}}]{PhysRevB.106.115139}%
  \BibitemOpen
  \bibfield  {author} {\bibinfo {author} {\bibfnamefont {E.}~\bibnamefont
  {Rosenberg}}, \bibinfo {author} {\bibfnamefont {J.~M.}\ \bibnamefont
  {DeStefano}}, \bibinfo {author} {\bibfnamefont {Y.}~\bibnamefont {Guo}},
  \bibinfo {author} {\bibfnamefont {J.~S.}\ \bibnamefont {Oh}}, \bibinfo
  {author} {\bibfnamefont {M.}~\bibnamefont {Hashimoto}}, \bibinfo {author}
  {\bibfnamefont {D.}~\bibnamefont {Lu}}, \bibinfo {author} {\bibfnamefont
  {R.~J.}\ \bibnamefont {Birgeneau}}, \bibinfo {author} {\bibfnamefont
  {Y.}~\bibnamefont {Lee}}, \bibinfo {author} {\bibfnamefont {L.}~\bibnamefont
  {Ke}}, \bibinfo {author} {\bibfnamefont {M.}~\bibnamefont {Yi}},\ and\
  \bibinfo {author} {\bibfnamefont {J.-H.}\ \bibnamefont {Chu}},\ }\href
  {https://doi.org/10.1103/PhysRevB.106.115139} {\bibfield  {journal} {\bibinfo
   {journal} {Phys. Rev. B}\ }\textbf {\bibinfo {volume} {106}},\ \bibinfo
  {pages} {115139} (\bibinfo {year} {2022})}\BibitemShut {NoStop}%
\bibitem [{\citenamefont {Pokharel}\ \emph {et~al.}(2022)\citenamefont
  {Pokharel}, \citenamefont {Ortiz}, \citenamefont {Chamorro}, \citenamefont
  {Sarte}, \citenamefont {Kautzsch}, \citenamefont {Wu}, \citenamefont {Ruff},\
  and\ \citenamefont {Wilson}}]{PhysRevMaterials.6.104202}%
  \BibitemOpen
  \bibfield  {author} {\bibinfo {author} {\bibfnamefont {G.}~\bibnamefont
  {Pokharel}}, \bibinfo {author} {\bibfnamefont {B.}~\bibnamefont {Ortiz}},
  \bibinfo {author} {\bibfnamefont {J.}~\bibnamefont {Chamorro}}, \bibinfo
  {author} {\bibfnamefont {P.}~\bibnamefont {Sarte}}, \bibinfo {author}
  {\bibfnamefont {L.}~\bibnamefont {Kautzsch}}, \bibinfo {author}
  {\bibfnamefont {G.}~\bibnamefont {Wu}}, \bibinfo {author} {\bibfnamefont
  {J.}~\bibnamefont {Ruff}},\ and\ \bibinfo {author} {\bibfnamefont {S.~D.}\
  \bibnamefont {Wilson}},\ }\href
  {https://doi.org/10.1103/PhysRevMaterials.6.104202} {\bibfield  {journal}
  {\bibinfo  {journal} {Phys. Rev. Mater.}\ }\textbf {\bibinfo {volume} {6}},\
  \bibinfo {pages} {104202} (\bibinfo {year} {2022})}\BibitemShut {NoStop}%
\bibitem [{\citenamefont {Huang}\ \emph {et~al.}(2023)\citenamefont {Huang},
  \citenamefont {Cui}, \citenamefont {Huang}, \citenamefont {Huo},
  \citenamefont {Liu}, \citenamefont {Li}, \citenamefont {Liang}, \citenamefont
  {Chen}, \citenamefont {Sun}, \citenamefont {Shen}, \citenamefont {Zhang},\
  and\ \citenamefont {Wang}}]{huang2023anisotropic}%
  \BibitemOpen
  \bibfield  {author} {\bibinfo {author} {\bibfnamefont {X.}~\bibnamefont
  {Huang}}, \bibinfo {author} {\bibfnamefont {Z.}~\bibnamefont {Cui}}, \bibinfo
  {author} {\bibfnamefont {C.}~\bibnamefont {Huang}}, \bibinfo {author}
  {\bibfnamefont {M.}~\bibnamefont {Huo}}, \bibinfo {author} {\bibfnamefont
  {H.}~\bibnamefont {Liu}}, \bibinfo {author} {\bibfnamefont {J.}~\bibnamefont
  {Li}}, \bibinfo {author} {\bibfnamefont {F.}~\bibnamefont {Liang}}, \bibinfo
  {author} {\bibfnamefont {L.}~\bibnamefont {Chen}}, \bibinfo {author}
  {\bibfnamefont {H.}~\bibnamefont {Sun}}, \bibinfo {author} {\bibfnamefont
  {B.}~\bibnamefont {Shen}}, \bibinfo {author} {\bibfnamefont {Y.}~\bibnamefont
  {Zhang}},\ and\ \bibinfo {author} {\bibfnamefont {M.}~\bibnamefont {Wang}},\
  }\href@noop {} {\bibinfo {title} {Anisotropic magnetism and electronic
  properties of the kagome metal smv6sn6}} (\bibinfo {year} {2023}),\ \Eprint
  {https://arxiv.org/abs/2303.00627} {arXiv:2303.00627 [cond-mat.str-el]}
  \BibitemShut {NoStop}%
\bibitem [{\citenamefont {Zhang}\ \emph {et~al.}(2022)\citenamefont {Zhang},
  \citenamefont {Liu}, \citenamefont {Cui}, \citenamefont {Guo}, \citenamefont
  {Wang}, \citenamefont {Shi}, \citenamefont {Zhang}, \citenamefont {Wang},
  \citenamefont {Dong}, \citenamefont {Sun}, \citenamefont {Dun},\ and\
  \citenamefont {Cheng}}]{PhysRevMaterials.6.105001}%
  \BibitemOpen
  \bibfield  {author} {\bibinfo {author} {\bibfnamefont {X.}~\bibnamefont
  {Zhang}}, \bibinfo {author} {\bibfnamefont {Z.}~\bibnamefont {Liu}}, \bibinfo
  {author} {\bibfnamefont {Q.}~\bibnamefont {Cui}}, \bibinfo {author}
  {\bibfnamefont {Q.}~\bibnamefont {Guo}}, \bibinfo {author} {\bibfnamefont
  {N.}~\bibnamefont {Wang}}, \bibinfo {author} {\bibfnamefont {L.}~\bibnamefont
  {Shi}}, \bibinfo {author} {\bibfnamefont {H.}~\bibnamefont {Zhang}}, \bibinfo
  {author} {\bibfnamefont {W.}~\bibnamefont {Wang}}, \bibinfo {author}
  {\bibfnamefont {X.}~\bibnamefont {Dong}}, \bibinfo {author} {\bibfnamefont
  {J.}~\bibnamefont {Sun}}, \bibinfo {author} {\bibfnamefont {Z.}~\bibnamefont
  {Dun}},\ and\ \bibinfo {author} {\bibfnamefont {J.}~\bibnamefont {Cheng}},\
  }\href {https://doi.org/10.1103/PhysRevMaterials.6.105001} {\bibfield
  {journal} {\bibinfo  {journal} {Phys. Rev. Mater.}\ }\textbf {\bibinfo
  {volume} {6}},\ \bibinfo {pages} {105001} (\bibinfo {year}
  {2022})}\BibitemShut {NoStop}%
\bibitem [{\citenamefont {Ishikawa}\ \emph {et~al.}(2021)\citenamefont
  {Ishikawa}, \citenamefont {Yajima}, \citenamefont {Kawamura}, \citenamefont
  {Mitamura},\ and\ \citenamefont {Kindo}}]{doi:10.7566/JPSJ.90.124704}%
  \BibitemOpen
  \bibfield  {author} {\bibinfo {author} {\bibfnamefont {H.}~\bibnamefont
  {Ishikawa}}, \bibinfo {author} {\bibfnamefont {T.}~\bibnamefont {Yajima}},
  \bibinfo {author} {\bibfnamefont {M.}~\bibnamefont {Kawamura}}, \bibinfo
  {author} {\bibfnamefont {H.}~\bibnamefont {Mitamura}},\ and\ \bibinfo
  {author} {\bibfnamefont {K.}~\bibnamefont {Kindo}},\ }\href
  {https://doi.org/10.7566/JPSJ.90.124704} {\bibfield  {journal} {\bibinfo
  {journal} {Journal of the Physical Society of Japan}\ }\textbf {\bibinfo
  {volume} {90}},\ \bibinfo {pages} {124704} (\bibinfo {year} {2021})},\
  \Eprint {https://arxiv.org/abs/https://doi.org/10.7566/JPSJ.90.124704}
  {https://doi.org/10.7566/JPSJ.90.124704} \BibitemShut {NoStop}%
\bibitem [{\citenamefont {Dally}\ \emph {et~al.}(2021)\citenamefont {Dally},
  \citenamefont {Lynn}, \citenamefont {Ghimire}, \citenamefont {Michel},
  \citenamefont {Siegfried},\ and\ \citenamefont
  {Mazin}}]{PhysRevB.103.094413}%
  \BibitemOpen
  \bibfield  {author} {\bibinfo {author} {\bibfnamefont {R.~L.}\ \bibnamefont
  {Dally}}, \bibinfo {author} {\bibfnamefont {J.~W.}\ \bibnamefont {Lynn}},
  \bibinfo {author} {\bibfnamefont {N.~J.}\ \bibnamefont {Ghimire}}, \bibinfo
  {author} {\bibfnamefont {D.}~\bibnamefont {Michel}}, \bibinfo {author}
  {\bibfnamefont {P.}~\bibnamefont {Siegfried}},\ and\ \bibinfo {author}
  {\bibfnamefont {I.~I.}\ \bibnamefont {Mazin}},\ }\href
  {https://doi.org/10.1103/PhysRevB.103.094413} {\bibfield  {journal} {\bibinfo
   {journal} {Phys. Rev. B}\ }\textbf {\bibinfo {volume} {103}},\ \bibinfo
  {pages} {094413} (\bibinfo {year} {2021})}\BibitemShut {NoStop}%
\bibitem [{\citenamefont {Wang}\ \emph {et~al.}(2021)\citenamefont {Wang},
  \citenamefont {Neubauer}, \citenamefont {Duan}, \citenamefont {Yin},
  \citenamefont {Fujitsu}, \citenamefont {Hosono}, \citenamefont {Ye},
  \citenamefont {Zhang}, \citenamefont {Chi}, \citenamefont {Krycka},
  \citenamefont {Lei},\ and\ \citenamefont {Dai}}]{PhysRevB.103.014416}%
  \BibitemOpen
  \bibfield  {author} {\bibinfo {author} {\bibfnamefont {Q.}~\bibnamefont
  {Wang}}, \bibinfo {author} {\bibfnamefont {K.~J.}\ \bibnamefont {Neubauer}},
  \bibinfo {author} {\bibfnamefont {C.}~\bibnamefont {Duan}}, \bibinfo {author}
  {\bibfnamefont {Q.}~\bibnamefont {Yin}}, \bibinfo {author} {\bibfnamefont
  {S.}~\bibnamefont {Fujitsu}}, \bibinfo {author} {\bibfnamefont
  {H.}~\bibnamefont {Hosono}}, \bibinfo {author} {\bibfnamefont
  {F.}~\bibnamefont {Ye}}, \bibinfo {author} {\bibfnamefont {R.}~\bibnamefont
  {Zhang}}, \bibinfo {author} {\bibfnamefont {S.}~\bibnamefont {Chi}}, \bibinfo
  {author} {\bibfnamefont {K.}~\bibnamefont {Krycka}}, \bibinfo {author}
  {\bibfnamefont {H.}~\bibnamefont {Lei}},\ and\ \bibinfo {author}
  {\bibfnamefont {P.}~\bibnamefont {Dai}},\ }\href
  {https://doi.org/10.1103/PhysRevB.103.014416} {\bibfield  {journal} {\bibinfo
   {journal} {Phys. Rev. B}\ }\textbf {\bibinfo {volume} {103}},\ \bibinfo
  {pages} {014416} (\bibinfo {year} {2021})}\BibitemShut {NoStop}%
\bibitem [{\citenamefont {Bouvier}\ \emph {et~al.}(1991)\citenamefont
  {Bouvier}, \citenamefont {Lethuillier},\ and\ \citenamefont
  {Schmitt}}]{PhysRevB.43.13137}%
  \BibitemOpen
  \bibfield  {author} {\bibinfo {author} {\bibfnamefont {M.}~\bibnamefont
  {Bouvier}}, \bibinfo {author} {\bibfnamefont {P.}~\bibnamefont
  {Lethuillier}},\ and\ \bibinfo {author} {\bibfnamefont {D.}~\bibnamefont
  {Schmitt}},\ }\href {https://doi.org/10.1103/PhysRevB.43.13137} {\bibfield
  {journal} {\bibinfo  {journal} {Phys. Rev. B}\ }\textbf {\bibinfo {volume}
  {43}},\ \bibinfo {pages} {13137} (\bibinfo {year} {1991})}\BibitemShut
  {NoStop}%
\bibitem [{Note1()}]{Note1}%
  \BibitemOpen
  \bibinfo {note} {See the Supplemental Material at [] for additional X-ray
  scattering and susceptibility measurements.}\BibitemShut {Stop}%
\bibitem [{\citenamefont {Rosenfeld}\ and\ \citenamefont
  {Mushnikov}(2008)}]{ROSENFELD20081898}%
  \BibitemOpen
  \bibfield  {author} {\bibinfo {author} {\bibfnamefont {E.}~\bibnamefont
  {Rosenfeld}}\ and\ \bibinfo {author} {\bibfnamefont {N.}~\bibnamefont
  {Mushnikov}},\ }\href
  {https://doi.org/https://doi.org/10.1016/j.physb.2007.10.220} {\bibfield
  {journal} {\bibinfo  {journal} {Physica B: Condensed Matter}\ }\textbf
  {\bibinfo {volume} {403}},\ \bibinfo {pages} {1898} (\bibinfo {year}
  {2008})}\BibitemShut {NoStop}%
\bibitem [{\citenamefont {Venturini}\ \emph {et~al.}(1996)\citenamefont
  {Venturini}, \citenamefont {Fruchart},\ and\ \citenamefont
  {Malaman}}]{VENTURINI1996102}%
  \BibitemOpen
  \bibfield  {author} {\bibinfo {author} {\bibfnamefont {G.}~\bibnamefont
  {Venturini}}, \bibinfo {author} {\bibfnamefont {D.}~\bibnamefont
  {Fruchart}},\ and\ \bibinfo {author} {\bibfnamefont {B.}~\bibnamefont
  {Malaman}},\ }\href
  {https://doi.org/https://doi.org/10.1016/0925-8388(95)01998-7} {\bibfield
  {journal} {\bibinfo  {journal} {Journal of Alloys and Compounds}\ }\textbf
  {\bibinfo {volume} {236}},\ \bibinfo {pages} {102} (\bibinfo {year}
  {1996})}\BibitemShut {NoStop}%
\bibitem [{\citenamefont {Penc}\ \emph {et~al.}(1999)\citenamefont {Penc},
  \citenamefont {Hofmann}, \citenamefont {Leciejewicz},\ and\ \citenamefont
  {Szytula}}]{B_Penc_1999}%
  \BibitemOpen
  \bibfield  {author} {\bibinfo {author} {\bibfnamefont {B.}~\bibnamefont
  {Penc}}, \bibinfo {author} {\bibfnamefont {M.}~\bibnamefont {Hofmann}},
  \bibinfo {author} {\bibfnamefont {J.}~\bibnamefont {Leciejewicz}},\ and\
  \bibinfo {author} {\bibfnamefont {A.}~\bibnamefont {Szytula}},\ }\href
  {https://doi.org/10.1088/0953-8984/11/39/313} {\bibfield  {journal} {\bibinfo
   {journal} {Journal of Physics: Condensed Matter}\ }\textbf {\bibinfo
  {volume} {11}},\ \bibinfo {pages} {7579} (\bibinfo {year}
  {1999})}\BibitemShut {NoStop}%
\bibitem [{\citenamefont {Blanco}\ \emph {et~al.}(1991)\citenamefont {Blanco},
  \citenamefont {Gignoux},\ and\ \citenamefont {Schmitt}}]{PhysRevB.43.13145}%
  \BibitemOpen
  \bibfield  {author} {\bibinfo {author} {\bibfnamefont {J.~A.}\ \bibnamefont
  {Blanco}}, \bibinfo {author} {\bibfnamefont {D.}~\bibnamefont {Gignoux}},\
  and\ \bibinfo {author} {\bibfnamefont {D.}~\bibnamefont {Schmitt}},\ }\href
  {https://doi.org/10.1103/PhysRevB.43.13145} {\bibfield  {journal} {\bibinfo
  {journal} {Phys. Rev. B}\ }\textbf {\bibinfo {volume} {43}},\ \bibinfo
  {pages} {13145} (\bibinfo {year} {1991})}\BibitemShut {NoStop}%
\bibitem [{\citenamefont {Rotter}\ \emph {et~al.}(2001)\citenamefont {Rotter},
  \citenamefont {Loewenhaupt}, \citenamefont {Doerr}, \citenamefont
  {Lindbaum},\ and\ \citenamefont {Michor}}]{PhysRevB.64.014402}%
  \BibitemOpen
  \bibfield  {author} {\bibinfo {author} {\bibfnamefont {M.}~\bibnamefont
  {Rotter}}, \bibinfo {author} {\bibfnamefont {M.}~\bibnamefont {Loewenhaupt}},
  \bibinfo {author} {\bibfnamefont {M.}~\bibnamefont {Doerr}}, \bibinfo
  {author} {\bibfnamefont {A.}~\bibnamefont {Lindbaum}},\ and\ \bibinfo
  {author} {\bibfnamefont {H.}~\bibnamefont {Michor}},\ }\href
  {https://doi.org/10.1103/PhysRevB.64.014402} {\bibfield  {journal} {\bibinfo
  {journal} {Phys. Rev. B}\ }\textbf {\bibinfo {volume} {64}},\ \bibinfo
  {pages} {014402} (\bibinfo {year} {2001})}\BibitemShut {NoStop}%
\end{thebibliography}%

\end{document}